\documentclass[shortnote,twocolumn]{jpsj2}
\usepackage{graphicx}

\def\BC{\bb C}
\def\_\BC{\bbi C}

\newcommand{\ev}{\equiv}
\newcommand{\lb}{\label}

\def\runtitle{A Matrix Model for 
Fractional Quantum Hall States}

\def\runauthor{A. \textsc{Jellal}, E.H. \textsc{Saidi} and  
H.B. \textsc{Geyer}}

\title{ A Matrix Model for a Class of  
Fractional Quantum Hall States
\thanks{To appear in J. Phys. Soc. Japan (2003)}}

\author{%
A. \textsc{Jellal}$^{1}$\thanks{{{jellal@gursey.gov.tr}}},
E.H. \textsc{Saidi}$^{2}$,
H.B. \textsc{Geyer}$^{3}$
and
R.A. \textsc{R\"omer}$^{4}$}

\inst{%
$^1$Institut f\"{u}r Physik Technische 
Universit\"{a}t, D-09107, Chemnitz, Germany \\
$^2$Lab/UFR, High Energy Physics, 
Physics Department, Mohammed V University,\\
Av. Ibn Battouta, P.O. Box. 1014, Rabat, Morocco\\
$^3$ Institute for Theoretical Physics, University of
Stellenbosch, Private Bag X1, \\ Matieland 7602, South Africa\\
$^4$ Department of Physics, University of Warwick, Coventry CV4 7AL, UK
}

\recdate{July 18, 2002}

\abst{
}
\kword{%
Non-commutative Chern-Simons, Matrix Model Theory,
Fractional Quantum Hall (FQH) Fluids 
}

\begin{document}
\sloppy
\maketitle

%%%%%%%%%%%%%%%%%%%%%%%%%%%%%%%%%%%%%%%%%%%%%%%%%%%%%%%%%%%%%%
\section{Introduction}
%%%%%%%%%%%%%%%%%%%%%%%%%%%%%%%%%%%%%%%%%%%%%%%%%%%%%%%%%%%%%%

%Laughlin~\cite{laughlin} showed that fractional quantum Hall effect
%at filling factor $\nu={1\ov 2m+1}$, where $m$ is integer,
%can be descibed by his famous trial wavefunctions:
%\begin{equation}
%\lb{lau}
%\Psi_{\rm L}^m = \prod_{i<j} (z_i-z_j)^{2m+1} e^{-{1\ov 4}\sum_i |z_i|^2}.
%\end{equation}

Recently, Susskind~\cite{susskind} showed that an Abelian
non-commutative Chern-Simons theory at level $k$ is actually
equivalent to Laughlin theory~\cite{laughlin}:
\begin{equation}
S=\frac{k}{4\pi} \int d^3y  \epsilon^{\mu \nu \lambda}
\left[
A_{\mu} \star \partial_{\nu} A_{\lambda}
+\frac{2}{3} A_{\mu} \star   A_{\nu} \star   A_{\lambda}
\right]
\end{equation}
where the star--product is the usual Moyal product with parameter
$\theta$. Therefore, he obtained the filling factor
\begin{equation} 
\nu_{\rm S}=\frac{1}{k}. 
\end{equation}

He also pointed out that the above theory can be formulated in terms
of a matrix model involving classical Hermitian matrix variables
$A_0,X^i$, $i=1,2$.  The Lagrangian for the matrix theory is
\begin{equation}
L=B \; {\rm{Tr}}
\left\{
\epsilon_{ij} (\dot{X}^i +i[A_0,X^i])X^j + 2\theta A_0
\right\}
\end{equation}
$B$ is the magnetic field. The equation of motion for the coordinate
$A_0$ (Gauss law constraint) is
\begin{equation}
[X^1,X^2] = i \theta
\end{equation}
which can only be solved if the matrices are infinite dimensional.
This corresponds to an infinite number of electrons on an infinite
plane.

For a finite system, Polychronakos~\cite{poly} has introduced an
additional set of bosonic degrees of freedom $\psi_m$, $m= 1,2,...,M$,
such that $\psi=(\psi_1,\cdots,\psi_M)$,
\begin{equation}
L_{\psi} = \psi^{\dag}(i \dot{\psi} -A_0 \psi).
\end{equation}
Considering $L+L_{\psi}$, Polychronakos~\cite{poly} found a quantum
correction to Susskinds filling factor such that
\begin{equation}
\lb{sp}
\nu_{\rm P}=\frac{1}{k+1}.
\end{equation}
In this case, the Gauss law constraint becomes
\begin{equation}
\lb{law}
[X^1,X^2]=i\theta
\left(
{{\bf 1}}- \frac{1}{k+1}\psi \psi^{\dag}
\right) .
\end{equation}

Later Hellerman and Van Raamsdonk~\cite{hellerman1} built the
corresponding wavefunctions for $L+L_{\psi}$,
\begin{equation}
|k\rangle =
\left\{
\epsilon^{i_1\cdots i_M}(\psi^{\dag})_{i_1}
(\psi^{\dag} A^{\dag})_{i_2}
...(\psi^{\dag} {A^{\dag}}^{M-1})_{i_M}
\right\}^k
|0\rangle
\end{equation}
where $\epsilon^{i_1\cdots i_M}$ is the fully antisymmetric tensor.
These are similar to Laughlins wavefunction~\cite{laughlin}.
Subsequently, three of us generalised~\cite{jellal1} the above results
to any filling factor given by
\begin{equation}
\nu _{k_{1} k_{2}}=\frac{1}{k_{1}}+\frac{1}{k_{2}},
\ \ \ \ \ k_{2}>k_{1}.
\end{equation}

In what follows, we propose a matrix model to describe such FQH states
that are not of Laughlin type.

\section{$\nu_{k_1k_2}$ fractional quantum Hall states}%-------------------

Although the $\nu =\frac{2}{5}$ FQH state is not of the Laughlin type,
it shares some basic features of Laughlin fluids. The point is that
from the standard definition of the filling factor $\nu
=\frac{N}{N_{\phi }}$, the state $\nu =\frac{2}{5}$ can naively be
thought of as corresponding to $\nu =\frac{N}{N_{\phi }}$ where the
number $N_{\phi }$ of flux quanta is given by a fractional amount of
the electron number; that is
\begin{equation}
N_{\phi }=(3-{\frac{1}{2}})N.
\end{equation}
In fact this way of viewing things reflects the original idea of a
hierarchical construction of FQH states for general filling factor
$\frac{p}{q}$.  In Haldane's hierarchy~\cite{haldane}, the elements of
the series
\begin{equation} 
\nu_{p_{1}p_{2}}=\frac{p_{2}}{p_{1}p_{2}-1}
\end{equation}
correspond to taking $N_{\phi }$ as given by a specific rational
factor of the electron number, i.e.,
\begin{equation} 
N_{\phi}=(p_{1}-\frac{1}{p_{2}})N.
\end{equation}
Upon setting 
\begin{equation}
\begin{array}{l}
k_{1}=p_{1},\ \ \ \ 
k_{2}=k_{1}(k_{1}p_{2}-1) \equiv r k_1
\end{array}
\end{equation}
we have
$\nu _{p_{1} p_{2}}\equiv \nu_{k_{1} k_{2}}$.
For $\nu =\frac{2}{5}$, e.g.,
\begin{equation}
\nu =\frac{2}{5}\equiv\frac{1}{3}+\frac{1}{15}.
\end{equation}

\section{Matrix model analysis}%-----------------------------

To describe FQH fluids at $\nu_{k_1k_2}$, we consider the following
action for a system of $N=N_1+N_2$ particles~\cite{jellal1}
\begin{equation}
\begin{array}{l}
\lb{ninact}
{\cal S} =\int dt\;
\sum_{i=1}^{2}\Big[ {\frac{k_{i}}{4\theta }}
{\rm{Tr}}\Big( i{ \bar{Z}_{i}}DZ_{i}-
\omega {\bar{Z}_{i}}Z_{i}+2\theta A_{0,i}\Big) 
\Big] \\ 
% \qquad 
\mbox{ } +h.c. +\int dt
\Big[\frac{i}{2}{\bar{\Psi}}^{{\alpha}{a}}
\left[\partial _{t}+{A_{0,1}}_{\alpha }^{{\beta}} 
\delta _{a}^{{b }}+{A_{0,2}}_{a}^{{b}} 
\delta _{\alpha }^{{\beta }}\right]  
\Psi _{\beta b}\\ 
%\qquad 
\mbox{ } 
+\lambda {\bar{\Psi}}^{{\alpha}{a}}
{Z_{1} }_{\alpha }^{{\beta}} Z_{2a}^{{b}}
\Psi_{\beta b}\Big]+ h.c.
\end{array}
\end{equation}
where $1\leq\alpha ,\beta \leq N_{1}$, $1\leq a,b\leq N_{2}$, $Z_l=
X_l^1 + i X_l^2$ and $A_{0,i}$ the gauge for the $i$th particle.  The
$J_{\alpha {\alpha}}^{(1)}$ and $J_{a {a}}^{(2)}$ currents (Gauss law
constraints) read as
\begin{equation}
\begin{array}{l}
\lb{nq4ncon4}
J_{\alpha {\alpha}}^{(1)}  =
 [Z_{1},{\bar{Z}_{1}}]_{\alpha {\alpha}}+
 \frac{\theta }{2k_{1}}\left( \sum_{a=1}^{N_{2}}
\Psi _{\alpha a}{\bar{\Psi}}_{ {\alpha} {a}}
-J_{0}^{(1)}\right) \\
J_{a {a}}^{(2)}  = 
[Z_{2},{\bar{Z}_{2}}]_{a {a}}+\frac{\theta }
{2k_{2}} \left( \sum_{\alpha =1}^{N_{1}}
\Psi _{\alpha a}{\bar{\Psi}}_{{\alpha}{a} }-
J_{0}^{(2)} \right)
\end{array}
\end{equation}
where the two $U(1)$ charge operators $J_{0}^{(1)}$ and $J_{0}^{(2)}$
are
\begin{equation}
\begin{array}{l}
\lb{chop}   
J_{0}^{(1)}  = J_{0}^{(2)}=J_{0}= \sum_{\alpha =1}^{N_{1}}
\sum_{a=1}^{N_{2}}\ \ 
{\bar{\Psi}}_{{\alpha}{a}} \Psi_{\alpha a}
\end{array} .
\end{equation}\\
\indent The wavefunctions $|\Phi \rangle $ describing the $(N_1+N_2)$
system of electrons on the non-commutative plane ${\mathbb R}_{\theta
  }^{2}$ with filling factor $\nu_{k_{1}k_{2}}$ should obey the
constraint~\cite{jellal1}
\begin{equation}
J_{0}|\Phi \rangle = \frac{N}{\nu_{k_{1}k_{2}}}
|\Phi \rangle.   
\end{equation}     
Once we know the fundamental state $|\Phi
_{\nu_{k_{1}k_{2}}}^{(0)}\rangle $, excitations are immediately
determined by applying the usual rules. Upon recalling the coordinate
operators as
\begin{equation}
Z_{1\alpha {{ \alpha}}}=
\sqrt{\frac{\theta }{2}}r_{\alpha {{\alpha}}}^{+},\ \ \ \
Z_{2a{ {a}}}=\sqrt{\frac{\theta }{2}} s_{a{a}}^{+}
\end{equation} 
the total Hamiltonian ${\cal H}$ may be treated as the sum of a free
part given by
\begin{equation}
\lb{ham1}
 {\cal H}_{0}=\frac{\omega}{2} 
\left(2 {\cal N}_{1}+ 2{\cal N}_{2 }+
N_{1}^{2}+N_{2}^{2} \right)
\end{equation}
where ${\cal N}_{1}= \sum_{\alpha ,\beta=1}^{N_{1}} r_{\alpha \beta
  }^{\dagger }r_{\beta \alpha }^{-}$ and ${\cal N} _{2 }=
\sum_{a,b=1}^{N_{2}}s_{ab}^{\dagger }s_{ba}^{-}$ are the operator
numbers counting the $N_1$ and $N_2$ particles respectively, and an
interacting part
\begin{equation}
\lb{ham2}
{\cal H}_{\rm int}\sim \left( {\psi }_{{a}{\alpha} }^{+}
r_{\alpha {{\beta }}}^{+}\ s_{a{{b}}}^{-}
\psi_{\beta b}^{-}+h.c.\right)
\end{equation}
describing couplings between the {\it two sectors}~\cite{jellal1}.
The creation and annihilation operators $r_{\alpha {{ \alpha}}}^{\pm
  }$, $s_{a{{a}}}^{\pm }$,\ and $\psi _{\alpha a}^{\pm }$ satisfy the
Heisenberg algebra
\begin{equation}
\begin{array}{l}
\lb{heisa}
\left[ \left( r^{-}\right) _{\alpha }^{{\alpha }},
\left(r^{+}\right) _{\beta }^{{\beta }}\right]  
\ \ \;\;= \delta _{\alpha \beta},  \ \ \ \ 
\left[ \left( s^{-}\right) _{a}^{{a}},
\left( s^{+}\right) _{b}^{{b}}\right]  
= \delta _{ab} \\
\left[ \left( \psi ^{-}\right) ^{{\alpha }
{a}},\left( \psi^{+}\right) _{\alpha a}\right]  
= 1  
\end{array}
\end{equation}
all others are given by commuting relations.

\section{Wavefunctions}%-----------------------------
A way to build the
spectrum of the Hamiltonian ${\cal H}_{0}$ is given by help of the
special condensate operators
\begin{equation}
\left( A^{+}\right)_{a\alpha }^{\left(n,m\right) }=
\Big[ \left( s^{+}\right) ^{n-1}\psi^{+}
\left( r^{+}\right)^{m-1} \Big] _{a\alpha }.
\end{equation}
The wavefunctions for the vacuum $|0>$ of ${\cal H}_0$ read as
\begin{equation}
\lb{wavef}
\left[ \varepsilon ^{\alpha _{1}...\alpha _{N_{1}}}
\prod_{j=1}^{p}O_{\alpha_{\left( jN_{2}+1\right) }...
\alpha _{\left( j+1\right) N_{2}}}^{\left(j\right) }
\right] ^{k_{1}}|0>
\end{equation}
where the $O^{\left( j\right) }$'s are building blocks and given by
\begin{equation}
\begin{array}{l}
\lb{builb}
O_{\alpha _{\left( jN_{2}+1\right) }...
\alpha _{\left( j+1\right)
N_{2}}}^{\left( j\right) }=
\varepsilon ^{a_{\left( jN_{2}+1\right)}...
a_{\left( j+1\right) N_{2}}} \\
\mbox{ }
\times 
\left( A^{+}\right) _{a_{\left(
jN_{2}+1\right) }
\alpha _{\left( jN_{2}+1\right) }}^{\left( 1,j\right)}
...
\left( A^{+}\right) _{a_{\left( j+1\right) N_{2}}
\alpha _{\left(
j+1\right) N_{2}}}^{\left( N_{2},j\right) }.
\end{array}
\end{equation}
%where 
%$\varepsilon ^{\alpha _{1}...\alpha _{N_{1}}}$ 
%is antisymmetric tensor.
The corresponding energy spectrum $E_{c}\left(\nu_{k_{1}k_{2}}\right)$
is
\begin{equation}
\begin{array}{l}
\lb{nener}
% E_{c}\left(\nu_{k_{1}k_{2}}\right)
E_{c}=
k_{1}\left[ p\frac{\left( N_{2}-1\right) \left(
N_{2}-2\right) }{2}+\frac{\left( p-1\right) 
\left( p-2\right) }{2}N_{2} \right] 
%\\ 
%\qquad\; \qquad\;\; 
+\frac{N_{1}+N_{2}}{2}.
\end{array}
\end{equation}
Note that for large value of $N_{1}$ and $N_{2}$ $(N_{1} = rN_{2})$,
$E_{c}(\nu_{k_{1}k_{2}})$ behaves quadratically in $N_2$,
\begin{equation}
E_{c}(\nu_{k_{1}k_{2}}) \sim \frac{k_2}{2}N_2^2.
\end{equation}
This energy relation is less than the total energy
$E_{d}(\nu_{k_{i}})$ of the decoupled configuration $\left( |\Phi
  _{1},v_{k_{1}} \rangle \otimes |\Phi _{2},v_{k_{2}}\rangle\right)$:
\begin{equation} 
E_{d}(\nu_{k_{i}})\ev E\left( \frac{1}{k_{1}}\right)
+E\left(\frac{1}{k_{2}}\right)
\sim
\frac{k_{2}(r+1)}{2}\ N_2^{2}.
\end{equation}
Therefore, we have the following relation
\begin{equation}
\lb{dif2}
E_d 
\sim \left( r+1\right) E_c.
\end{equation}

For the example of the FQH state at $\nu =\frac{2}{5}$, the energy of
the decoupled representation reads as
\begin{equation}
\lb{decap}
E\left(\frac{1}{3}\right)+E\left(\frac{1}{15}\right)\sim 45 N_2^{2}
\end{equation}
while that of the interacting one is 
\begin{equation}
\lb{cap}
E_{c}\left(\frac{2}{5}\right)\sim \frac{15}{2}N_2^{2}
\end{equation}
leading to
\begin{equation}
E\left(\frac{1}{3}\right)+E\left(\frac{1}{15}\right)
\sim 6 E_{c}\left(\frac{2}{5}\right).
\end{equation}

\section{Conclusion}%--------------------------------------

We have developed a matrix model for FQH states at filling factor
$\nu_{k_1k_2}$ going beyond the Laughlin theory. To illustrate our
idea, we have considered an FQH system of a finite number
$N=\left(N_{1}+N_{2}\right) $ of electrons with filling factor
$\nu_{k_{1}k_{2}}\ev \nu_{p_{1}p_{2}}=\frac{p_{2}}{p_{1}p_{2}-1}$; $\ 
p_{1}$ is an odd integer and $p_{2}$ is an even integer.  The
$\nu_{p_{1}p_{2}}$ series corresponds just to the level two of the
Haldane hierarchy; it recovers the Laughlin series $\nu_{p_{1}}
=\frac{1}{p_{1}}$ by going to the limit $\ p_{2}$ large and contains
several observable FQH states such as $\nu = \frac{2}{3}, 
\frac{2}{5},
\cdots$.

%\section*{Acknowledgment}
%It is a great pleasure to acknowledge
%the organizers of 
%{\it
%Kolloquium zum DFG-Schwerpunktprogramm Quanten-Hall-Systeme,
%9-10 January (2003)}, Physikzentrum, Bad Honnef-Germany,
%for their kind invitation.

\end{document}